\documentclass[12pt,a4paper]{article}

\usepackage{amsmath}

\usepackage{amssymb}
\usepackage{amsfonts}
\usepackage{graphicx}
\usepackage{subfigure}
\usepackage{color}
\usepackage{amsmath}
\usepackage{amsthm}

\newcommand{\bx}{\boldsymbol{x}}

\newcommand{\by}{\boldsymbol{y}}

\newcommand{\bw}{\boldsymbol{w}}

\newcommand{\argmax}{\operatornamewithlimits{argmax}}
\newcommand{\argmin}{\operatornamewithlimits{argmin}}

\theoremstyle{definition}

\newtheorem{theorem}{Theorem}[section]
\newtheorem{corollary}{Corollary}[section]
\newtheorem{lemma}{Lemma}[section]

\title{Maximum-a-posteriori estimation with Bayesian confidence regions}
\author{Marcelo Pereyra \footnote{School of Mathematics, University of Bristol, Bristol BS8 1TW, United Kingdom (marcelopereyra@ieee.org).}}

\begin{document}

\maketitle

\begin{abstract}
Solutions to inverse problems that are ill-conditioned or ill-posed may have significant intrinsic uncertainty. Unfortunately, analysing and quantifying this uncertainty is very challenging, particularly in high-dimensional problems. As a result, while most modern mathematical imaging methods produce impressive point estimation results, they are generally unable to quantify the uncertainty in the solutions delivered. This paper presents a new general methodology for approximating Bayesian high-posterior-density credibility regions in inverse problems that are convex and potentially very high-dimensional. The approximations are derived by using recent concentration of measure results related to information theory for log-concave random vectors. A remarkable property of the approximations is that they can be computed very efficiently, even in large-scale problems, by using standard convex optimisation techniques. In particular, they are available as a by-product in problems solved by maximum-a-posteriori estimation. The approximations also have favourable theoretical properties, namely they outer-bound the true high-posterior-density credibility regions, and they are stable with respect to model dimension. The proposed methodology is illustrated on two high-dimensional imaging inverse problems related to tomographic reconstruction and sparse deconvolution, where the approximations are used to perform Bayesian hypothesis tests and explore the uncertainty about the solutions, and where proximal Markov chain Monte Carlo algorithms are used as benchmark to compute exact credible regions and measure the approximation error.
\end{abstract}
%


%
\section{Introduction}
\label{sec:intro}

Recovering an unobserved image from raw noisy data is a central topic in imaging sciences, especially for images that are only observed partially or with limited resolution. Canonical examples include, for instance, image denoising \cite{Lebrun:2013,Aharon:2015}, deconvolution \cite{Bonettini2013,Molina2013}, compressive sensing \cite{Donoho2006,Candes2006}, super-resolution \cite{Babacan:2011,Veniamin2016}, tomographic reconstruction \cite{Bioucas:2007,Lustig:2007}, inpainting \cite{Chan2011,Mairal:2008}, source separation \cite{VCA:2005,Zhengming:2012}, fusion \cite{JBD02015,Haro:2015}, and phase retrieval \cite{Candes:2013,PUMA:2007}. These image recovery or estimation tasks often require solving a high-dimensional inverse problem that is ill-posed or ill-conditioned, and that may consequently involve significant uncertainty about the unobserved true image \cite{uncertaintyBook}. Following intensive research efforts, the last decade has witnessed tremendous advances in methodology for imaging inverse problems, with most methods now adopting formal approaches to derive solutions and to study the underpinning algorithms. Particularly, convex inverse problems have received a lot of attention in the late, leading to important developments in theory, methods, models, and algorithms for this class of problems.

There are many formal mathematical frameworks available to address imaging inverse problems \cite{somersalo:2005}. In particular, many modern methods use the Bayesian statistical framework; that is, they use statistical models to represent the data observation process and the prior knowledge available, and they derive solutions by using Bayesian inference techniques \cite{somersalo:2005}. Especially, maximum-a-posteriori (MAP) estimation has been adopted as a standard approach for high-dimensional imaging problems, particularly for convex problems where MAP estimates can be computed efficiently by using large-scale convex optimisation algorithms \cite{Green2015} .

Despite the great progress in estimation accuracy and computing time, there are still some fundamental open problems in imaging sciences that limit its impact. In particular, most modern methodologies produce accurate point estimates but are unable to quantify the uncertainty in the solutions delivered. Uncertainty quantification is important in many applications related to quantitative imaging, scientific inquiry, and decision-making, where it is necessary to analyse images as high-dimensional physical measurements and not as pictures.

Following a Bayesian uncertainty quantification approach, this paper presents a general method for computing approximate joint posterior credible regions (i.e., Bayesian confidence regions) for inverse problems that are convex and potentially very high-dimensional. These approximations will enable exploring the uncertainty about the solutions, for example by performing Bayesian hypothesis tests. A key property is that the approximations can be computed efficiently by convex optimisation, and that they are available as a by-product of MAP estimation. 

The remainder of the paper is organised as follows: section \ref{sec:BayesianConfidence} introduces some elements of Bayesian analysis that are essential to our method and specifies the class of inverse problems considered. In section \ref{sec:method} we present the proposed method for approximating Bayesian high-posterior-density credibility regions and analyse its theoretical properties. Section \ref{sec:experiments} illustrates the method on two high-dimensional imaging inverse problems related to tomographic reconstruction and sparse deconvolution, where the approximations are used to perform Bayesian hypothesis tests and explore the uncertainty about the solutions. Conclusions and perspectives for future work are finally reported in section \ref{sec:conclusion}.


\section{Bayesian uncertainty quantification}
\label{sec:BayesianConfidence}
Let $\bx \in \mathbb{R}^n$ be an unknown signal of interest, and $\by$ an observation related to $\bx$ by a statistical model with likelihood function $p(\by|\bx)$. Suppose that the recovery of $\bx$ from $\by$ is ill-posed or ill-conditioned, resulting in significant uncertainty about the true value of $\bx$ \cite{somersalo:2005}. Bayesian inference methods address this difficulty by using prior knowledge about $\bx$ to reduce the uncertainty and deliver accurate estimation results \cite{somersalo:2005}. Precisely, they model $\bx$ as a random vector with prior distribution $p(\bx)$ promoting expected structural or regularity properties (e.g., sparsity or smoothness), and combine observed and prior information by using Bayes' theorem, leading to the posterior distribution \cite{cprbayes}
$$
p(\bx|\by) = \frac{p(\by|\bx)p(\bx)}{\int_{\mathbb{R}^n} p(\by|\bx)p(\bx)\textrm{d}\bx}\,,
$$
which models our knowledge about $\bx$ after observing $\by$. In this paper we assume that $p(\bx|\by)$ is log-concave, i.e.,
\begin{eqnarray}\label{posterior}
p(\bx|\by) = \exp{\{-g_{\by}(\bx)\}} / Z_{\by}\,,
\end{eqnarray}
where $g_{\by}(\bx)$ is a convex function and $Z_{\by} \in \mathbb{R}$ is a normalising constant that is possibly unknown. Notice that the class \eqref{posterior} comprises many important models that are used extensively in data science, particularly models of the form $g_{\by}(\bx) = \|\by-A\bx\|^2 /2\sigma^2 + \phi(B\bx) + \boldsymbol{1}_\mathcal{S}(\bx)$ for some linear operators $A$, $B$, convex regulariser $\phi$, and convex set constraint $\mathcal{S}$.

When $\bx$ is high-dimensional, drawing conclusions directly from $p(\bx|\by)$ is not possible. As a result, we use summaries of $p(\bx|\by)$, particularly point estimators, that capture some of the information about $\bx$ that is relevant for the application considered \cite{cprbayes}. High-dimensional inference methods typically use the MAP estimator of $\bx$, i.e., 
\begin{eqnarray}\label{map}
\begin{split}
\hat{\bx}_{MAP} = \argmax_{\bx \in \mathbb{R}^n} p(\bx|\by) = \argmin_{\bx \in \mathbb{R}^n} g_{\by}(\bx),
\end{split}
\end{eqnarray}
that can often be computed efficiently by convex optimisation \cite{boydbook, pesquet:2011, Parikh:2014} (as opposed other summaries and estimators that generally require high-dimensional integration w.r.t. $p(\bx|\by)$ \cite{Pereyra2016}).

However, in its raw form MAP estimation fails to deliver some basic aspects of the statistical inference paradigm \cite{Green2015}. In particular, given the uncertainty that is inherent to ill-posed and ill-conditioned inverse problems, it would be highly desirable to not only deliver point estimates such as $\hat{\bx}_{MAP}$, but also posterior credibility sets that indicate the region of the parameter space where most of the posterior probability mass of $\bx$ lies. This is formalised in the Bayesian decision theory framework by computing \emph{credible regions} \cite{cprbayes}. A set $C_\alpha$ is a posterior credible region with confidence level $(1-\alpha)\%$ if
$$
\textrm{P} \left[\bx \in C_{\alpha} | \by \right] = 1-\alpha,
$$ 
where it is recalled that $\textrm{P} \left[\bx \in C_{\alpha} | \by \right] = \int p(\bx|\by)\boldsymbol{1}_{C_{\alpha} }(\bx)\textrm{d}\bx$. It is easy to check that for any $\alpha \in (0,1)$ there are infinitely many regions of the parameter space that verify this property. Here we consider the so-called \emph{highest posterior density} (HPD) region, which is decision-theoretically optimal in the sense that it has minimum volume \cite{cprbayes}, and is given by 
\begin{eqnarray}\label{HPD}
C^*_{\alpha} = \{ \bx : g_{\by}(\bx) \leq \gamma_\alpha \}
\end{eqnarray}
with $\gamma_\alpha \in \mathbb{R}$ chosen such that $\int_{C^*_\alpha} p(\bx|\by) \textrm{d}\bx = 1-\alpha$ holds. In addition to being optimal in this sense, this joint credible set has the important advantage that it can be enumerated by simply specifying the scalar value $\gamma_\alpha$ (whereas enumerating an arbitrary convex set in $\mathbb{R}^n$ remains an open problem). 

Unfortunately, computing credible sets is very challenging when $n$ is large because it requires calculating integrals of the form $\int_{\mathbb{R}^n} \boldsymbol{1}_{C_\alpha} (\bx) p(\bx|\by) \textrm{d}\bx$. These integrals can be approximated with high accuracy by Monte Carlo integration \cite{robert:casella98,Pereyra2016} (for instance by using the state-of-the-art proximal Markov chain Monte Carlo algorithm \cite{Pereyra2015}). However, the computational cost related to approximating integrals is often several orders of magnitude higher than that involved in optimisation for MAP estimation, and it increases rapidly with problem dimension \cite{Pereyra2016,Green2015}. As a result, most high-dimensional inference methods do not quantify uncertainty.

\section{Approximating HPD regions by convex optimisation}\label{sec:method}
\subsection{Proposed approximation}
The main contribution of this paper is to exploit the log-concavity of $p(\bx|\by)$ to derive a conservative approximate confidence region $\tilde{C}_{\alpha}$ that contains (i.e., outer-bounds) the true HPD region $C^*_{\alpha}$, and which has the fundamental advantage of being straightforward to compute by using modern convex optimisation algorithms, even in very high dimensions. In particular, we propose an approximation whose computation only assumes knowledge of the MAP estimator $\hat{\bx}_{MAP}$, and does not require evaluating expectations nor the normalising constant $Z_{\by} =\int_{\mathbb{R}^n} \exp{\{-g_{\by}(\bx)\}} \textrm{d}\bx$ which often become computationally intractable as $n \rightarrow \infty$.

\begin{theorem}\label{Theo1}
Suppose that the posterior distribution $p(\bx|\by) = \exp{\{-g_{\by}(\bx)\}}/Z_{\by}$ is log-concave on $\mathbb{R}^n$. Then, for any $\alpha \in (4\exp{(-n/3)},1)$, the highest-posterior-density region $C^*_{\alpha}$ is contained within the bounding set 
$$
\tilde{C}_{\alpha} = \{ \bx : g_{\by}(\bx) \leq  g_{\by}(\hat{\bx}_{MAP}) + n(\tau_\alpha+1) \},
$$
with positive constant $\tau_\alpha = \sqrt{16\log(3/\alpha)/n}$ independent of $p(\bx|\by)$, and where\\ $\hat{\bx}_{MAP} = \argmin_{\bx \in \mathbb{R}^n} g_{\by}(\bx)$ is the maximum-a-posteriori estimator of $\bx$ given $\by$.

\end{theorem}
\noindent{\textit{Proof.}} To prove Theorem \ref{Theo1} we use two recent results from information theory. The first result is a probability concentration inequality recently proposed in \cite{bobkov2011}, which for the purpose of our proof we write in the following form
\begin{lemma}\label{Lemma1}
Suppose that $p(\bx|\by) = \exp{\{-g_{\by}(\bx)\}}/Z_{\by}$ is log-concave on $\mathbb{R}^n$, then
$$
\textrm{P} \left[|g_{\by}(x) - \textrm{E}\{ g_{\by}(\bx)\}| \geq \tau n \right] \leq 3 \exp{(-\tau^2 n /16)}, 
$$
for any $\tau \in [0,2]$, and where the expectation $\textrm{E}\{ g_{\by}(\bx)\} = \int_{\mathbb{R}^n} g_{\by}(\bx)p(\bx|\by) \textrm{d}\bx$.
\end{lemma}
This result follows directly from \cite[Theorem 1.2]{bobkov2011} by setting $t = \tau^2 n$ and noting that $\textrm{E}\{\log p(\bx|\by)\} - \log p(\bx|\by) = g_{\by}(x) - \textrm{E}\{ g_{\by}(\bx)\}$)
\noindent Lemma \ref{Lemma1} is related to a concentration property of log-concave random vectors: as $n$ grows the probability mass of $\bx$ concentrates on a typical set on the neighbourhood of the $(n-1)$-dimensional shell $\{\bx : g_{\by}(\bx) = \textrm{E}\{ g_{\by}(\bx)\}\}$. Lemma \ref{Lemma1} implies that for $\tau \in [0,2]$, the probability $\textrm{P} \left[g_{\by}(x) \geq \textrm{E}\{ g_{\by}(\bx)\} + \tau n \right] \leq 3 \exp{(-\tau^2 n /16)}$. To derive an upper bound for the confidence level $(1-\alpha)$ we set $\tau_\alpha = \sqrt{16\log(3/\alpha)/n}$ and obtain the inequality\\ $\textrm{P} \left[g_{\by}(x) > \textrm{E}\{ g_{\by}(\bx)\} + \tau_\alpha n \right] \leq \alpha$. Following on from this, to construct a bound that does not require computing the (generally computationally intractable) expectation $\textrm{E}\{ g_{\by}(\bx)\}$, we use Proposition I.2 of \cite{Bobkov2011b} to derive the inequality
\begin{eqnarray}\label{inequality1}
\textrm{E}\{g_{\by}(\bx)\} \leq  g_{\by}(\hat{\bx}_{MAP}) + n,
\end{eqnarray}
which holds for all log-concave distributions on $\mathbb{R}^n$. The proof is then concluded by using this result to show that 
\begin{eqnarray}\label{inequality2}
\textrm{P} \left[\bx \in \tilde{C}_{\alpha} \biggr | \by \right] \geq 1-\alpha,
\end{eqnarray}
where $\tilde{C}_{\alpha} = \{ \bx : g_{\by}(\bx) \leq g_{\by}(\hat{\bx}_{MAP}) + n(\tau_\alpha+1)\}$, and where it is easy to check that $g_{\by}(\hat{\bx}_{MAP}) + n(\tau_\alpha+1) \geq \gamma_\alpha$ and therefore by construction $C^*_{\alpha} \subseteq \tilde{C}_{\alpha}$. \qed

\subsection{Approximation error analysis}
Theorem \ref{Theo1} essentially states that $\tilde{C}_{\alpha}$ is a conservative approximation of $C^*_{\alpha}$, with the important computational advantage that it is available as a by-product in any convex problem that is solved by MAP estimation. Following on from this, a natural questions is whether $\tilde{C}_{\alpha}$ is an accurate approximation of $C^*_{\alpha}$, particularly in high-dimensional settings. To study this question we analyse the error involved in approximating $\gamma_\alpha$, the true threshold value of the HPD region $C^{*}_{\alpha}$, with the surrogate threshold $\tilde{\gamma}_\alpha = g(\hat{\bx}_{MAP}) + n(\tau_\alpha+1)$ associated with the approximation $\tilde{C}_{\alpha}$. We first derive a general non-asymptotic bound for finite $n$ and then consider asymptotic bounds for distributions with specific tail behaviours.


\begin{theorem}\label{Theo2}
Suppose that the posterior distribution $p(\bx|\by) = \exp{\{-g_{\by}(\bx)\}}/Z_{\by}$ is log-concave on $\mathbb{R}^n$, then
$$
0 \leq \tilde{\gamma}_\alpha - \gamma_ \alpha  \leq  \eta_\alpha\sqrt{n} + n,
$$
with positive constant $\eta_\alpha = \sqrt{16\log(3/\alpha)} + \sqrt{1/\alpha}$ independent of $p(\bx|\by)$.

\noindent{\textit{Proof.}} To prove Theorem \ref{Theo2} we construct the inequality
\begin{eqnarray}\label{inequality3}
\begin{split}
\tilde{\gamma}_\alpha - \gamma_ \alpha \leq |\tilde{\gamma}_\alpha - \textrm{E}\{g_{\by}(\bx)\}| + |\gamma_ \alpha - \textrm{E}\{g_{\by}(\bx)\}|,
\end{split}
\end{eqnarray}
and derive upper bounds for each term; the lower bound $\tilde{\gamma}_\alpha - \gamma_ \alpha \geq 0$ follows from the fact that $\tilde{C}_{\alpha}$ is a conservative approximation of $C^*_{\alpha}$. To upper bound the first term of \eqref{inequality3} we use \eqref{inequality1} to establish that $\textrm{E}\{g_{\by}(\bx)\} \in [g_{\by}(\hat{\bx}_{MAP}), g_{\by}(\hat{\bx}_{MAP})+n]$ and derive the inequality
\begin{eqnarray}\label{inequality4}
\begin{split}
|\tilde{\gamma}_\alpha - \textrm{E}\{g_{\by}(\bx)\}| \leq n(\tau_\alpha+1),
\end{split}
\end{eqnarray}
where we have used the definition $\tilde{\gamma}_\alpha = g_{\by}(\hat{\bx}_{MAP}) + n(\tau_\alpha+1)$. To upper bound the second term of \eqref{inequality3} we use the fact that because $p(\bx|\by)$ is log-concave then \cite{nguyen2013}
\begin{eqnarray}\label{varIneq}
\textrm{Var}\{g_{\by}(\bx)\} = \textrm{E}\{g_{\by}(\bx)^2\} - \textrm{E}\{g_{\by}(\bx)\}^2 \leq n.
\end{eqnarray}
From Chebyshev's inequality, for all $\zeta >0$
\begin{equation*}
P \left(|g_{\by}(\bx) - \textrm{E}\{g_{\by}(\bx)\}| > \zeta \right) \leq \textrm{Var}\{g_{\by}(\bx)\} /\zeta^2.
\end{equation*}
Then, using \eqref{varIneq} and setting $\zeta = |\gamma_ \alpha - \textrm{E}\{g_{\by}(\bx)\}|$, we obtain that
\begin{eqnarray}\label{inequality5}
P \left(|g_{\by}(\bx) - \textrm{E}\{g_{\by}(\bx)\}| > |\gamma_ \alpha - \textrm{E}\{g_{\by}(\bx)\}| \right) \leq n(|\gamma_ \alpha - \textrm{E}\{g_{\by}(\bx)\})^{-2}.
\end{eqnarray}
Moreover, $P[g_{\by}(\bx) > \gamma_ \alpha] = \alpha$ by construction of $C^*_{\alpha}$, which implies that
\begin{eqnarray}\label{inequality6}
P \left(|g_{\by}(\bx) - \textrm{E}\{g_{\by}(\bx)\}| > |\gamma_ \alpha - \textrm{E}\{g_{\by}(\bx)\}| \right) \geq \alpha.
\end{eqnarray}
Finally, inequalities \eqref{inequality5} and \eqref{inequality6} imply that
$|\gamma_ \alpha - \textrm{E}\{g_{\by}(\bx)\}| \leq \sqrt{n/\alpha}$,
and together with \eqref{inequality4} that $\tilde{\gamma}_\alpha - \gamma_ \alpha  \leq  (\sqrt{16\log(3/\alpha)} + \sqrt{1/\alpha})\sqrt{n} + n$ concluding the proof. \qed

%
%
\end{theorem}

Theorem \ref{Theo2} leads to two interesting observations about the approximation $\tilde{C}_{\alpha}$. First, $\tilde{C}_{\alpha}$ is a stable approximation, as the error $\tilde{\gamma}_\alpha - \gamma_ \alpha$ grows at most linearly with $n$ when $n$ is large. Second, $\tilde{C}_{\alpha}$ is asymptotically tight for the class of log-concave distributions, in the sense that the normalised error $(\tilde{\gamma}_\alpha - \gamma_ \alpha)/n \rightarrow 0$ as $n \rightarrow \infty$. To establish this point, as well as to develop an intuition about the relationship between the approximation error and the shape of the tails of $p(\bx|\by)$ (which determine the shape of ${C}^*_{\alpha}$ and $\tilde{C}_{\alpha}$), we consider the following sequence of log-concave distributions:
 \begin{corollary}\label{Coro2}
Let $\mathbb{X} = \{x_n, n \in \mathbb{N}\}$ be discrete-time stochastic process that takes values in $\mathbb{R}$. Suppose that for each $n \in \mathbb{N}$ the random vector $\bx^{(n)} = (x_1,\cdots,x_n)$ has marginal distribution $p_n(\bx^{(n)}) = \exp{\{-\lambda \sum_{i=1}^n |x_i |^q \}}/\lambda^{-n/q}$ with $q \in [1, \infty)$ and $\lambda \in \mathbb{R}^+$, then
$$
\lim_{n\rightarrow\infty} (\tilde{\gamma}^{(n)}_\alpha - \gamma^{(n)}_ \alpha)/n = 1 - 1/q,
$$
where, for each $n \in \mathbb{N}$, $\gamma_ \alpha^{(n)}$ and $\tilde{\gamma}_\alpha^{(n)}$ are respectively the threshold values of the HPD region $C^{*(n)}_{\alpha}$ and the approximation $\tilde{C}^{(n)}_{\alpha}$ associated with $p_n(\bx^{(n)})$.
\end{corollary}



The proof of Corollary \ref{Coro2} follows directly from the proof of Theorem \ref{Theo2}, and by using the fact that for distributions of the form $p_n(\bx^{(n)}) = \exp{\{-\lambda \sum_{i=1}^n |x_i |^q \}} /\lambda^{-n}$ with $q \in [1, \infty)$ and $\lambda \in \mathbb{R}^+$, $\lim_{n\rightarrow \infty} |\textrm{E}\{\log p_n(\bx^{(n)})\}+ n(\tau^{(n)}_\alpha+1) -\log p_n(\hat{\bx}_{MAP}^{(n)}) |/n = 1- 1/q$. 

Notice from Corollary \ref{Coro2} that the approximation error vanishes for $q = 1$ and $n \rightarrow \infty$, therefore the lower bound of Theorem \ref{Theo2} is asymptotically tight and $\tilde{C}_{\alpha}$ is exact in this case. Similarly, the upper bound of Theorem \ref{Theo2} is also asymptotically tight since it is attained when $n \rightarrow \infty$ and $q \rightarrow \infty$ (therefore this upper bound cannot be improved without constraining the tails of the log-concave distributions considered). Finally, it is worth mentioning that by proceeding in a similar fashion to Corollary \ref{Coro2}, it can be shown that if $\mathbb{X}$ is a stationary ergodic Gaussian process then $\lim_{n\rightarrow\infty} (\tilde{\gamma}^{(n)}_\alpha - \gamma^{(n)}_ \alpha)/n = 0.5$.

\begin{figure}[htbp!]
\begin{minipage}[l2]{0.49\linewidth}
  \centering
  \centerline{\includegraphics[width=7.5cm]{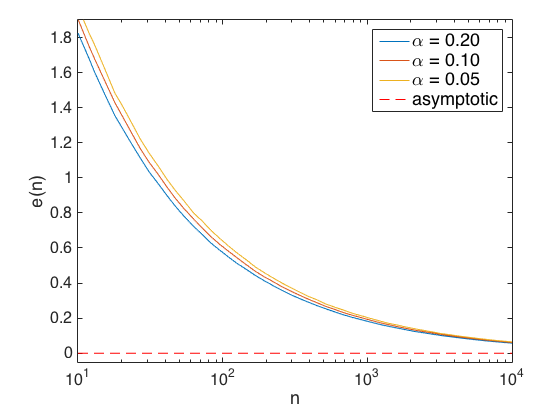}}
  \small{(a) Laplace experiment}
\end{minipage}
\begin{minipage}[l2]{0.49\linewidth}
  \centering
  \centerline{\includegraphics[width=7.5cm]{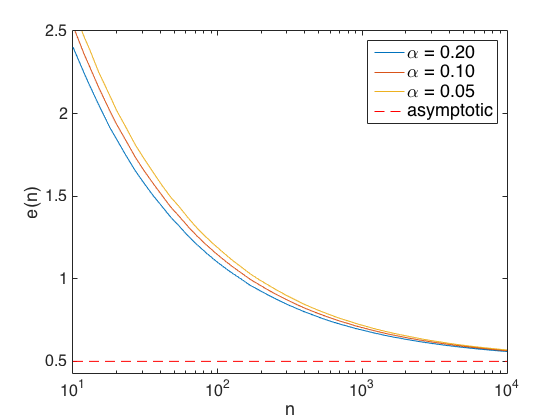}}
  \small{(b) Gaussian experiment}
\end{minipage}
\caption{\small{Comparison between the approximation error $e(n) = (\tilde{\gamma}^{(n)}_\alpha - \gamma^{(n)}_ \alpha)/n$ calculated by Monte Carlo integration and the asymptotic error given by Corollary \ref{Coro2} for $n = 1,\ldots, 10^4$, $\alpha = 0.2, 0.1, 0.05$, and the following two models: (a) Laplace $p_n(\bx^{(n)}) = \exp{\{- \sum_{i=1}^n |x_i |\}}$, and (b) Gaussian $p_n(\bx^{(n)}) = \exp{\{- \sum_{i=1}^n x_i ^2 \}}$.}} \label{asymptoticError}
\end{figure}

Furthermore, to assess the scale of values of $n$ for which these asymptotic results come into effect, Figure \eqref{asymptoticError} compares the approximation error $e(n) = (\tilde{\gamma}^{(n)}_\alpha - \gamma^{(n)}_ \alpha)/n$ calculated by Monte Carlo integration with the asymptotic error given by Corollary \ref{Coro2}. Figure \eqref{asymptoticError} (a) shows the true and asymptotic approximation errors as a function of $n$ for a Laplace distribution ($q = 1$) and for $\alpha = 0.2$, $\alpha = 0.1$ and $\alpha = 0.05$ (the asymptotic error is depicted in a dashed red). Similarly, Figure \eqref{asymptoticError} (b) shows the true and asymptotic errors for a Gaussian distribution ($q = 2$) and the same values of $\alpha$. We observe that in both cases, and for all the values of $\alpha$ considered, the approximation error falls sharply as $n$ increases, with the asymptotics clearly coming into effect for $n > 10^3$.

\subsection{Connections with approximate Bayesian inference approaches}
{We conclude this section with a discussion of alternative approaches to perform approximate Bayesian inference efficiently. We focus on the main high-dimensional approaches in the literature: variational Bayesian methods, belief propagation algorithms, and analytic approximations.}

{Variational Bayesian computation methods deliver approximate inferences efficiently by replacing the posterior distribution with a tractable approximation, which is obtained by specifying a family of approximations with favourable tractability properties and selecting the element of the family that is closest to the true posterior in the sense of the Kullback-Leibler divergence \cite{Green2015}. The approximations used in the literature typically factorise as products of low-dimensional marginals, or are based on Gaussian distributions with specific covariance structures \cite{Pereyra2016,seeger,opper}. This approach is model-specific, in the sense that the approximations and inference algorithms are tailed for specific models. Over the past decade the approach has been applied successfully to several Bayesian models related to mathematical imaging, often delivering accurate point estimation results \cite{Pereyra2016}. Of course, the approximations can also be used to derive approximate credible sets such as HPD regions. Interestingly, unlike $\tilde{C}_\alpha$ which is conservative, credible regions obtained from variational approximations are typically liberal and underestimate posterior uncertainty. This is in part related to the fact that most approximations used in the literature have a sparser dependence structure than the true posterior (either because they factorise explicitly as products of univariate or low-dimensional densities, or implicitly because they rely on Gaussian approximations with limited covariance structures \cite{seeger,opper}). Because the performance of variational Bayesian methods depends strongly on the model and approximations considered, the resulting approximate credible regions are accurate for some models and perform poorly for others. Unfortunately, it is generally difficult to assess their accuracy (notice that variational approximations may simultaneously deliver accurate minimum mean-squared-error (MMSE) point estimation results if the mean of the approximating distribution is close to the true posterior mean, and poor credible regions due to approximation errors at the tails).}

{Belief propagation algorithms are an increasingly popular approach to perform approximate Bayesian inference \cite{Pereyra2016}. The algorithms are useful for approximating marginal densities and performing MMSE inference, and have been applied successfully to many linear inverse problems (particularly in the context of compressive sensing and of the so-called \emph{approximate-message-passing} algorithms) \cite{Pereyra2016}. In addition to point estimates, the algorithms can also be used to compute approximate marginal confidence intervals for each image pixel. Using a Bonferroni correction, an approximate $m$-dimensional joint credibility region with level $(1-\alpha)$ can then be obtained by collecting $m$ pixel-wise marginal intervals of level $(1-\alpha/m)$ and constructing an $m$-dimensional hyperrectangle \cite{miller}. The resulting credible regions are accurate when $m$ is very small (e.g. $m < 10$), but suffer from a course of dimensionality and deteriorate very quickly as $m$ increases (precisely, the corrections that enable constructing joint sets from from marginal intervals become very conservative as $m$ increases, which in turn leads to a dramatic loss in hypothesis testing power). Consequently, such uncertainty quantification methods have limited applicability in mathematical imaging, where $m$ is often large.}

{Finally, a third approach is to approximate $p(\bx|\by)$ analytically. The predominant approximation of this kind is the Gaussian approximation with mean $\hat{\bx}_{MAP}$ and precision matrix given by the Hessian of $g_y$ \cite{mackay} (this technique is closely related to variational Bayes Gaussian methods \cite{opper}). These approximations are widely used in applied statistics. Unfortunately, Gaussian approximations may perform poorly in imaging inverse problems for the following reasons: the approximations rely on a second order approximation of $g_y$ that may be inaccurate if $g_{\by}$ is not sufficiently smooth around $\hat{\bx}_{MAP}$; ill-posed and ill-conditioned problems involving non-identifiable or poorly-identifiable likelihoods, and non-Gaussian priors, typically exhibit strongly non-Gaussian tails; non-Gaussian behaviour also arises naturally in high-dimensional models involving parameter space constraints, particularly if $\hat{\bx}_{MAP}$ lies at the boundary \cite{bochkina2014}. Another drawback is that computations involving the Hessian matrix of $g_y$ often scale poorly with $n$. Nevertheless, it is worth mentioning that a new Bernstein-von Mises theorem might lead to interesting developments in this topic \cite{bochkina2014}. Precisely, it has been recently established that under certain conditions the non-Gaussian components of $\bx|\by$ converge to a gamma distribution as the dimension of $\by$ goes to infinity. This suggests a new type of analytic approximation combining Gaussian and gamma components (see \cite{bochkina2014} for an application to emission tomographic imaging). To the best of our knowledge, the properties of the credible regions derived from these new approximations have not been studied yet.}

{In conclusion, the main strengths of the proposed methodology are its generality, theoretical underpinning, and simplicity of application in problems solved by MAP estimation. Alternative approaches based on variational Bayesian and belief propagation algorithms are more model-specific. They are potentially very accurate for some models, but their accuracy for uncertainty quantification is often difficult to assess. Gaussian approximations are not generally well adapted to imaging inverse problems, though new approximations combining Gaussian and gamma components might be potentially significantly better for some models.}

\section{Experimental results}
\label{sec:experiments}
In this section we illustrate the proposed methodology with two canonical imaging inverse problems: tomographic image reconstruction with a total-variation prior, and sparse image deconvolution with an $\ell_1$ prior. In the Bayesian setting these problems are predominately solved by MAP estimation, making the computation of $\tilde{C}_\alpha$ straightforward. Here we use $\tilde{C}_\alpha$ to explore the posterior uncertainty about $\bx$ and analyse specific aspects about the solutions delivered, particularly by using $\tilde{C}_\alpha$ to conduct hypothesis tests. Moreover, to assess the approximation error we also compute exact HPD credibility regions ${C}^*_\alpha$ for each problem by Monte Carlo integration (we use the proximal Metropolis-adjusted Langevin algorithm \cite{Pereyra2015}, which is a state-of-the-art Markov chain Monte Carlo method specifically designed for high-dimensional distributions that are log-concave). All experiments were conducted on a Apple Macbook Pro computer running MATLAB 2015.

\subsection{Tomographic image reconstruction with total-variation prior}\label{tomographic_imaging}
We consider a tomographic image reconstruction problem with a total-variation prior. In this inverse problem the goal is to recover a high-resolution image $\bx \in \mathbb{R}^n$ from an incomplete and noisy set of Fourier measurements $\by \in \mathbb{C}^n$ related to $\bx$ by $\by = H F \bx + \bw$, where $F$ is the discrete Fourier transform operator, $H$ is a subsampling mask related to tomographic imaging, and $\bw \sim \mathcal{N}(0,\sigma^2\boldsymbol{I}_n)$. This problem is ill-posed, a difficulty that Bayesian methods address by exploiting prior knowledge about $\bx$. Here we use a prior based on the total-variation norm of $\bx$, which is widely used for this type of problem. The resulting posterior density is log-concave and is given by
\begin{eqnarray}\label{tomographic}
p(\bx|\by) \propto \exp{\left[-(\|\by-H F \bx\|^2/2\sigma^2 +\lambda \|\nabla_d\bx\|_{1-2})\right]},
\end{eqnarray}
where $\|\cdot\|_{1-2}$ is the composite $\ell_1 -\ell_2$ norm, $\nabla_d$ is the two-dimensional discrete gradient operator. As mentioned previously, Bayesian image reconstruction is predominantly solved by MAP estimation, and there are several convex optimisation algorithms that can be used to compute the maximiser of \eqref{tomographic} (here we use the ADMM algorithm SALSA \cite{Figueiredo2011}).

\begin{figure}[htbp!]
\begin{minipage}[l2]{0.49\linewidth}
  \centering
  \centerline{\includegraphics[width=7.5cm]{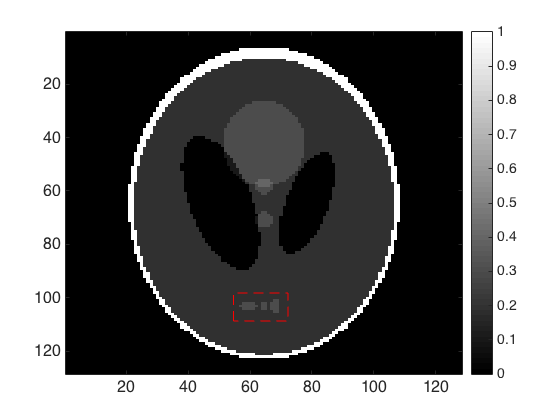}}
  \small{(a)}
\end{minipage}
\begin{minipage}[l2]{0.49\linewidth}
  \centering
  \centerline{\includegraphics[width=7.5cm]{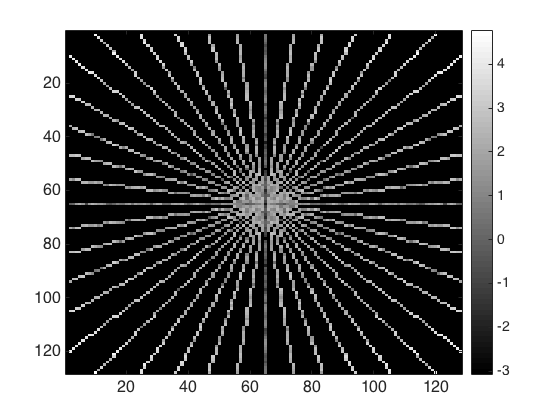}}
  \small{(b)}
\end{minipage}
\begin{minipage}[l2]{0.49\linewidth}
  \centering
  \centerline{\includegraphics[width=7.5cm]{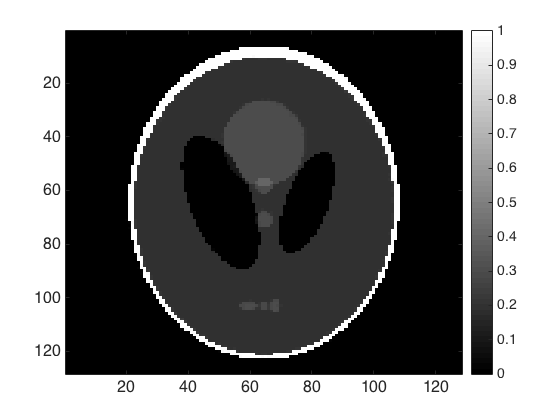}}
  \small{(c)}
\end{minipage}
\begin{minipage}[l2]{0.49\linewidth}
  \centering
  \centerline{\includegraphics[width=7.5cm]{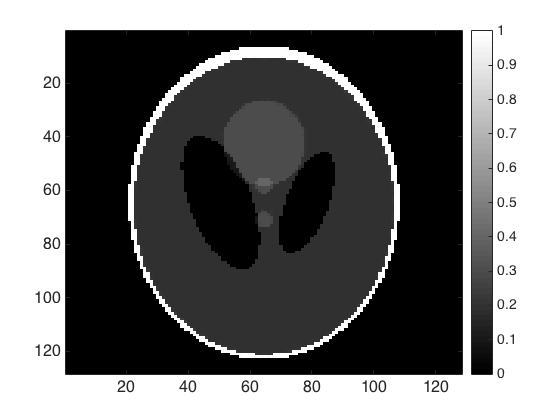}}
  \small{(d)}
\end{minipage}
\begin{minipage}[l2]{0.49\linewidth}
  \centering
  \centerline{\includegraphics[width=7.5cm]{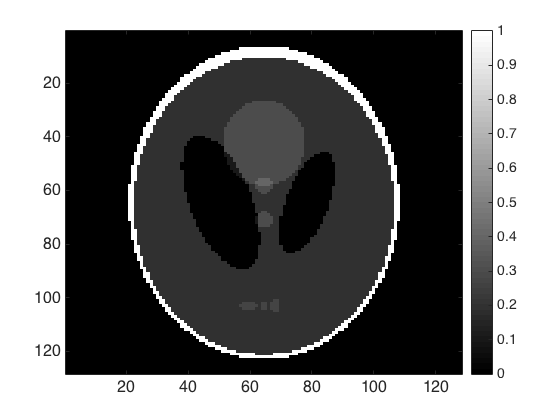}}
  \small{(e)}
\end{minipage}
\begin{minipage}[l2]{0.49\linewidth}
  \centering
  \centerline{\includegraphics[width=7.5cm]{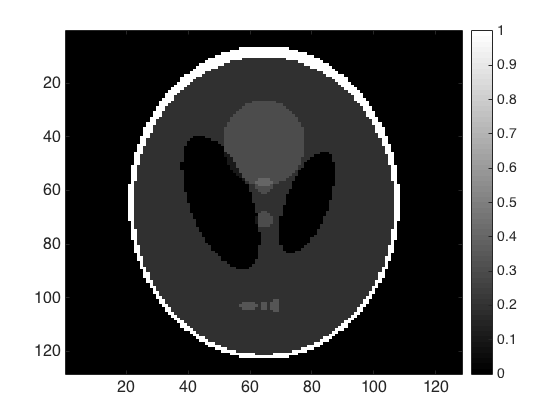}}
  \small{(f)}
\end{minipage}
\caption{\small{MRI experiment (high SNR): (a) \texttt{Shepp-Logan phantom} image ($128\times128$ pixels), (b) tomographic observation $\by$ (amplitude of Fourier coefficients in logarithmic scale, $\sigma = 7 \times 10^{-3}$), (c) MAP estimate $\hat{\bx}_{MAP}$ (the intensity of the structure of interest is $0.30$, the surrounding background intensity is $0.20$), (d) knockout test surrogate image $\bx_\dagger$, (e) surrogate $\hat{\bx}_{min}$ (the lower bound on structure intensity values is $0.27$), (f) surrogate $\hat{\bx}_{max}$ (the upper bound on structure intensity values is $0.33$)}} \label{FigMRI1}
\end{figure}

Figure \ref{FigMRI1} presents an experiment with the \texttt{Shepp-Logan phantom} magnetic resonance image (MRI) of size $n = 128 \times 128$ pixels displayed in Figure \ref{FigMRI1}(a). Figure \ref{FigMRI1}(b) shows a noisy tomographic measurement $\by$ of this image, generated using Gaussian noise with $\sigma = 7 \times 10^{-3}$ (to improve visibility Figure \ref{FigMRI1}(b) shows the amplitude of the Fourier coefficients in logarithmic scale, with black regions representing unobserved coefficients). Notice from Figure \ref{FigMRI1}(b) that only $15\%$ of the original Fourier coefficients are observed, suggesting potentially significant intrinsic uncertainty about the true image. Moreover, Figure \ref{FigMRI1}(c) shows the Bayesian estimate $\hat{\bx}_{MAP}$ associated with \eqref{tomographic} {(to compute this estimate we used the hyper-parameter value $\lambda = 180$, which we selected manually to obtain good reconstruction results; the automatic selection of $\lambda$ and its impact on uncertainty quantification are discussed in Section \ref{sec:conclusion})}. Computing this estimate with SALSA \cite{Figueiredo2011} required $0.75$ seconds. As expected, we observe that $\hat{\bx}_{MAP}$ provides an accurate estimation of the original image, confirming the good performance of the approach.

To illustrate the proposed method, we now focus on the structure highlighted in red in Figure \ref{FigMRI1}(c). Suppose that this structure is relevant from a clinical viewpoint because it provides important information for diagnosis or treatment related decision-making. Also suppose that we first observe this structure in the Bayesian estimate $\hat{\bx}_{MAP}$ and that, following on from this, we wish to explore the posterior uncertainty about $\bx$ to learn more about the structure and inform decisions. In this example, we first assess the evidence supporting that the structure is indeed present in true image (as opposed to being a noise artefact for example), and then examine the range of likely intensity values for its pixels. Precisely, the estimate $\hat{\bx}_{MAP}$ indicates that the intensity of the 3 bright spots is approximately $0.3$, compared to a surrounding background intensity is approximately $0.2$, and we seek to quantify the uncertainty about these values. 

To perform the first analysis we propose the following point hypothesis test based on a knockout approach: First, use $\hat{\bx}_{MAP}$ to compute $\tilde{C}_\alpha$. Second, we generate a surrogate image $\bx_\dagger$ by copying $\hat{\bx}_{MAP}$ and removing the structure of interest in a way that is as compatible with the prior distribution as possible (e.g., we apply a segmentation-inpainting process to replace the 3 bright spots with the surrounding intensity level). Third, we seek to reject the hypothesis that $\bx_\dagger$ belongs to the credible set $\tilde{C}_\alpha$ for a suitable confidence level $(1-\alpha)$ (e.g., $95\%$ or $99\%$). If $\bx_\dagger \notin \tilde{C}_\alpha$ the model rejects this hypothesis with confidence $(1-\alpha)$, suggesting that the structure is present in the true image with high probability. Conversely, if $\bx_\dagger \in \tilde{C}_\alpha$ the model fails to reject the hypothesis, indicating insufficient evidence for the structure considered. The rationale for this procedure is the following: $\bx_\dagger$ is obtained by modifying $\hat{\bx}_{MAP}$, which represents the centre of $\tilde{C}_\alpha$, by removing one specific feature of $\hat{\bx}_{MAP}$ in a way that most favourable to the prior distribution. If this single modification produces a solution $\bx_\dagger$ that is outside $\tilde{C}_\alpha$ this indicates that there is strong evidence in the likelihood for that specific feature. On the other hand, if $\bx_\dagger \in \tilde{C}_\alpha$ we conclude that the posterior uncertainty about this feature is to high to draw strong conclusions. {Note that this procedure generally overestimates uncertainty because it uses an $n$-dimensional credible region to explore properties of a subset of pixels of dimension $m < n$. Statistically more accurate results could be obtained by operating directly with the marginal posterior of interest, however this density is typically computationally intractable (see Section \ref{sec:conclusion} for more details)}.

By applying this procedure we obtain the surrogate image displayed in Figure \ref{FigMRI1}(d), which scores $g_{\by}(\bx_\dagger) = 2.91 \times 10^5$. This value is larger than the threshold $g_{\by}(\hat{\bx}_{MAP}) + n(\tau_\alpha+1) = 1.53 \times 10^5$ (we used $\alpha = 0.01$ related to a $99\%$ confidence level). Therefore $\bx_\dagger \notin \tilde{C}_\alpha$, rejecting the knockout hypothesis and providing evidence in favour or the structure considered (performing this test required $75$ milliseconds). 

Following on from this, we take our analysis further and assess the range of intensity values that this structure is likely to take in the true image. Precisely, to quantify the uncertainty about this intensity we generate two new surrogate test images, where we artificially increase and decrease the structure pixel values until the surrogates exit $\tilde{C}_{0.01}$. Figures \ref{FigMRI1}(e)-(f) show the limit solutions $\hat{\bx}_{min}$ and  $\hat{\bx}_{max}$ related to the minimum and maximum values that fall within $\tilde{C}_{0.01}$. These minimum and maximum intensity values are $0.27$ and $0.33$, indicating that the values of the order of $0.30$ reported in $\hat{\bx}_{MAP}$ have a level of uncertainty of approximately $10\%$ {(using the exact HPD credibility region $C^*_{0.01}$ computed with the Monte Carlo algorithm \cite{Pereyra2015} leads to the values $0.272$ and $0.327$, indicating an approximation error of order $1\%$)}.  

\begin{figure}[htbp!]
\begin{minipage}[l2]{0.49\linewidth}
  \centering
  \centerline{\includegraphics[width=7.5cm]{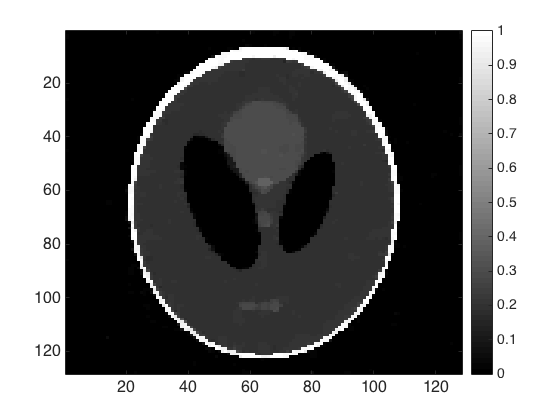}}
  \small{(a)}
\end{minipage}
\begin{minipage}[l2]{0.49\linewidth}
  \centering
  \centerline{\includegraphics[width=7.5cm]{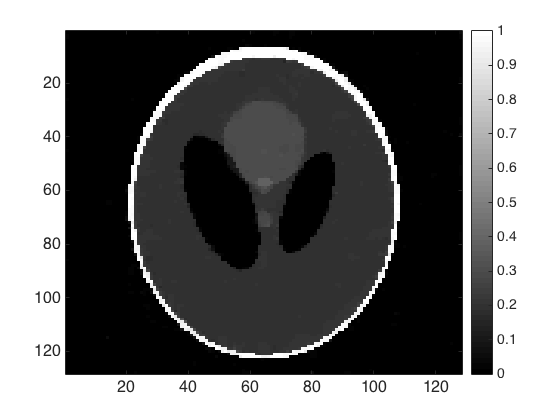}}
  \small{(b)}
\end{minipage}
\caption{\small{MRI experiment (low SNR): (a) MAP estimate $\hat{\bx}^\prime_{MAP}$, (b) knockout test surrogate image $\bx^\prime_\dagger$,.}} \label{FigMRI2}
\end{figure}

Furthermore, to illustrate how increasing the level of noise increases the posterior uncertainty about $\bx$, we repeated the experiment with a new observation $\by^\prime$ with worse signal-to-noise ratio (SNR), generated by using $\sigma = 7 \times 10^{-2}$ (recall that the previous observation was generated using $\sigma = 7 \times 10^{-3}$). Figures \ref{FigMRI2}(a)-(b) show respectively the new MAP estimate $\hat{\bx}^\prime_{MAP}$ and surrogate test image $\bx^\prime_\dagger$ obtained by repeating the approach described above with this new observation. In this case we obtain that $g_{\by}(\bx^\prime_\dagger) = 1.27 \times 10^4$, which is significantly lower than the threshold $g_{\by}(\hat{\bx}_{MAP}) + n(\tau_\alpha+1) = 2.85 \times 10^4$ (we used $\alpha = 0.2$ related to a mild confidence level of $80\%$). Therefore $\bx^\prime_\dagger \in \tilde{C}_\alpha$ and it is not possible to reject $\bx^\prime_\dagger$ as solution to the inverse problem (performing this test required $50$ milliseconds). We conclude that in this case, because of the lower SNR, it is not possible to assert confidently that the structure is present in the image. 

\begin{figure}[htbp!]
\begin{minipage}[l2]{0.49\linewidth}
  \centering
  \centerline{\includegraphics[width=7.5cm]{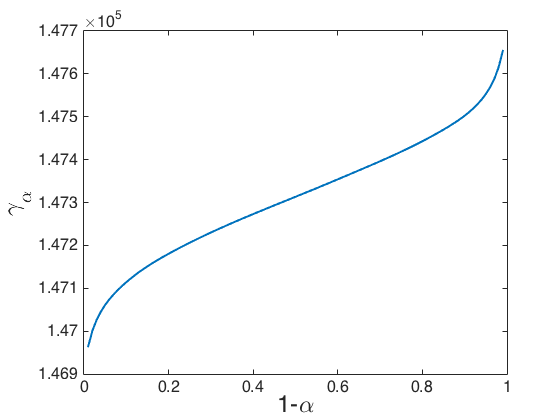}}
  \small{(a)}
\end{minipage}
\begin{minipage}[l2]{0.49\linewidth}
  \centering
  \centerline{\includegraphics[width=7.5cm]{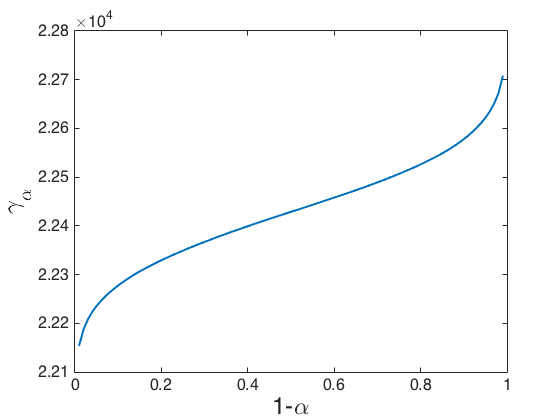}}
  \small{(b)}
\end{minipage}
\begin{minipage}[l2]{0.49\linewidth}
  \centering
  \centerline{\includegraphics[width=7.5cm]{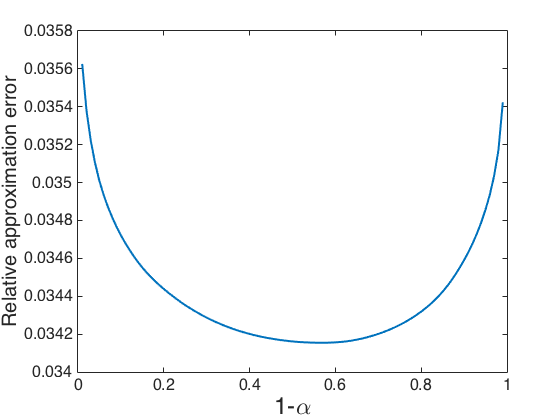}}
  \small{(a)}
\end{minipage}
\begin{minipage}[l2]{0.49\linewidth}
  \centering
  \centerline{\includegraphics[width=7.5cm]{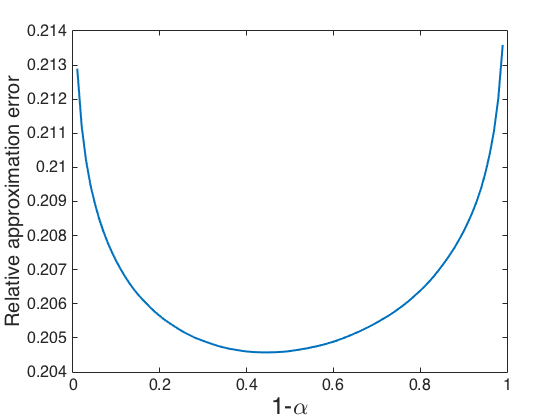}}
  \small{(b)}
\end{minipage}
\caption{\small{MRI experiment: (a) HDP region thresholds $\gamma_\alpha$ (high SNR), (b) HDP region thresholds $\gamma_\alpha$ (low SNR), (c) relative approximation error $(\tilde{\gamma}_\alpha-\gamma_\alpha)/\gamma_\alpha$ (high SNR), (d) relative approximation error $(\tilde{\gamma}_\alpha-\gamma_\alpha)/\gamma_\alpha$ (low SNR).}} \label{FigMRI3}
\end{figure}

Finally, we conclude this experiment by analysing the approximation errors related to using $\tilde{C}_{\alpha}$ instead of the exact HPD credibility region $C^*_\alpha$. Precisely, we used the Monte Carlo algorithm \cite{Pereyra2015} to compute  the regions $C^*_\alpha$ for both problems (reconstruction with high and low SNR), and for one hundred values of $\alpha \in (0,1)$. Figures \ref{FigMRI3}(a)-(b) show the values of the exact thresholds $\gamma_\alpha$ for each problem (computing these thresholds by Monte Carlo integration required 40 hours). Notice that for both models the difference between $\gamma_{0.01}$ and $\gamma_{0.99}$ is very small, confirming the intuition behind Lemma \ref{Lemma1} that the posterior probability mass is highly concentrated around an $(n-1)$-dimensional shell. Moreover, Figures \ref{FigMRI3}(c)-(d) report the relative error $(\tilde{\gamma}_\alpha - \gamma_\alpha)/\gamma_\alpha$ for each problem. We observe that the approximation errors are only of the order of $3\%$ and $20\%$, which is remarkably low given that $\tilde{C}_{\alpha}$ is guaranteed to outer-bound ${C}^*_{\alpha}$ for the class of log-concave distributions (hence it is generally not tight for specific models and datasets), and that the approximation is available as a by-product of MAP estimation with minimum computational cost.

\subsection{Sparse image deconvolution with $\ell_1$ prior}
The second experiment we consider is a non-blind Bayesian sparse image deconvolution problem with a Laplace or $\ell_1$ prior. In this canonical inverse problem the goal is to recover a high-resolution image $\bx \in \mathbb{R}^n$ from a known blurred and noisy observation $\by \in \mathbb{R}^n$ related to $\bx$ by $\by = H\bx + \bw$, where $H$ is a blurring operator and $\bw \sim \mathcal{N}(0,\sigma^2\boldsymbol{I}_n)$. Similarly to \ref{tomographic_imaging}, this inverse problem is ill-posed, a difficulty that Bayesian image deconvolution methods address by exploiting prior knowledge about $\bx$. Here we use a Laplace prior related to the $\ell_1$ norm of $\bx$, which is widely used for this type of problem. The resulting posterior density is log-concave and is given by
\begin{eqnarray}\label{deconvolution}
p(\bx|\by) \propto \exp{\left[-(\|\by-H\bx\|^2/2\sigma^2 +\lambda \|\bx\|_{1})\right]}.
\end{eqnarray}
As mentioned previously, Bayesian image deconvolution is predominantly solved by MAP estimation, and there are several convex optimisation algorithms to compute the maximiser of \eqref{deconvolution} (here we use the ADMM algorithm SALSA \cite{Figueiredo2011}).

\begin{figure}[htbp!]
\begin{minipage}[l2]{0.49\linewidth}
  \centering
  \centerline{\includegraphics[width=7.5cm]{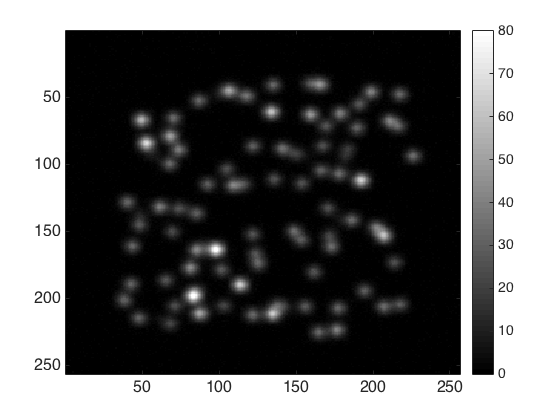}}
  \small{(a)}
\end{minipage}
\begin{minipage}[l2]{0.49\linewidth}
  \centering
  \centerline{\includegraphics[width=7.5cm]{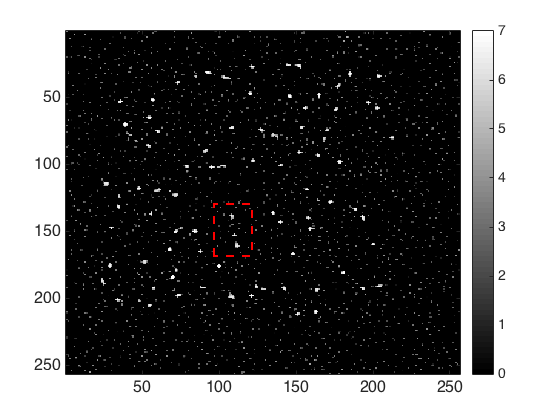}}
  \small{(b)}
\end{minipage}
\begin{minipage}[l2]{0.49\linewidth}
  \centering
  \centerline{\includegraphics[width=7.5cm]{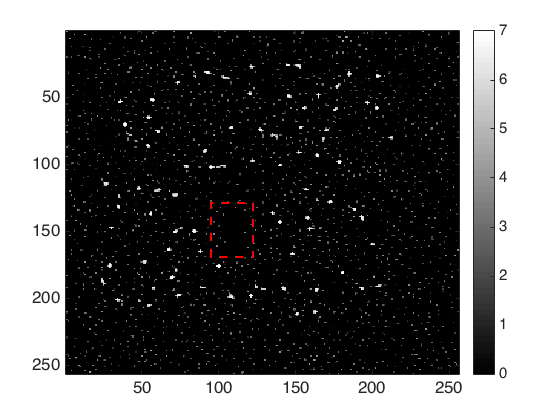}}
  \small{(c)}
\end{minipage}
\begin{minipage}[l2]{0.49\linewidth}
  \centering
  \centerline{\includegraphics[width=7.5cm]{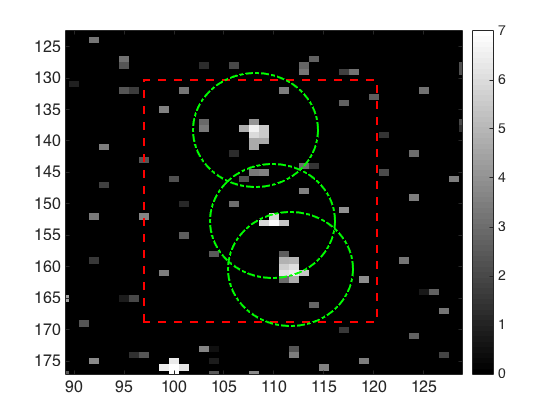}}
  \small{(d)}
\end{minipage}
\caption{\small{Microscopy experiment: (a) Blurred image $\by$ ($256\times256$ pixels, $4\mu m \times 4\mu m$)),\newline (b) MAP estimate $\bx_{MAP}$ (logarithmic scale), (c) knockout test surrogate image $\bx_\dagger$, \newline (d) molecule position uncertainty quantification (vertical: $\pm 93nm$, horizontal $\pm 140nm$).}} \label{FigMicro}
\end{figure}

Figure \ref{FigMicro} presents an experiment with a microscopy dataset of \cite{Zhu2012} related to high-resolution live cell imaging. Figure \ref{FigMicro}(a) shows an observation $\by$ of a sample of $100$ molecules over a field of size $4 \mu m \times 4 \mu m$, acquired with an application specific point-spread-function of size $16 \times 16$ pixels and a blurred signal-to-noise ratio of $20$dB (see \cite{Zhu2012} for more details). Figure \ref{FigMicro}(b) shows the Bayesian estimate $\hat{\bx}_{MAP}$ associated with \eqref{deconvolution} (notice that $\hat{\bx}_{MAP}$ is displayed in logarithmic scale to improve visibility - {to compute this estimate we used the hyper-parameter value $\lambda = 0.01$, which we selected manually to obtain good deconvolution results; the automatic selection of $\lambda$ and its impact on uncertainty quantification are discussed in Section \ref{sec:conclusion}}). Computing this estimate with SALSA \cite{Figueiredo2011} required $2.3$ seconds. Notice from Figure \ref{FigMicro}(b) that the deconvolution process has restored the fine detail in the image, allowing a better identification of the molecules.

To illustrate the proposed method, we now focus on the specific group of molecules in the region of interested highlighted in red (see Figure \ref{FigMicro}(b)). Suppose that these specific moles are relevant for the application considered and that, after observing them in $\hat{\bx}_{MAP}$, we wish to assess the uncertainty about their exact position. In a manner akin to Section \ref{tomographic_imaging} we first conduct a knockout test to check that the molecules are present in the true image (e.g., as opposed to being an artefact due to noise), and then perform additional tests to determine the uncertainty about their position. 

To conduct the knockout test we first create a surrogate image $\bx_\dagger$ by copying $\hat{\bx}_{MAP}$ and removing the molecules of interest such that the resulting image is as favourable to the prior distribution as possible. The resulting test image is displayed in Figure \ref{FigMicro}(c). Second, we check if $\bx_\dagger$ belongs to the credible region $\tilde{C}_\alpha$ (we use $\alpha = 0.01$ related to a $99\%$ confidence level) and obtain that $\bx_\dagger \notin \tilde{C}_\alpha$, suggesting that the group of molecules considered is present in the true image with high probability (precisely, we obtain that $g(\bx_\dagger) = 1.19 \times 10^5$, which is larger than the $\tilde{C}_\alpha$ threshold $g_{\by}(\hat{\bx}_{MAP}) + n(\tau_\alpha+1) = 1.03 \times 10^5$). Following on from this, we explore $\tilde{C}_\alpha$ to quantify the uncertainty about the exact position of the molecules. Precisely, we generate a collection of new surrogate test images by copying $\hat{\bx}_{MAP}$ and displacing molecules in different directions until the surrogates exit $\tilde{C}_\alpha$. Figure \ref{FigMicro}(d) depicts a focus on the region of interest, with the posterior uncertainty of the molecule positions shown in dashed green. According to this analysis the uncertainty at level $99\%$ is of the order of $\pm 6$ pixels vertically and $\pm 9$ pixels horizontally, corresponding to $\pm 93\textrm{nm}$ and $\pm 140\textrm{nm}$ {(using the exact HPD credibility region $C^*_{0.01}$ computed with the Monte Carlo algorithm \cite{Pereyra2015} leads to the values $\pm 78\textrm{nm}$ and $\pm 125\textrm{nm}$, indicating an approximation error of order $15\textrm{nm}$ or $1$ pixel)}. It is worth mentioning that these results are in close in agreement with the experimental precision results reported in \cite{Zhu2012}, which identified an average precision of the order of $80\textrm{nm}$ for the one hundred molecules. 

\begin{figure}[htbp!]
\begin{minipage}[l2]{0.49\linewidth}
  \centering
  \centerline{\includegraphics[width=7.5cm]{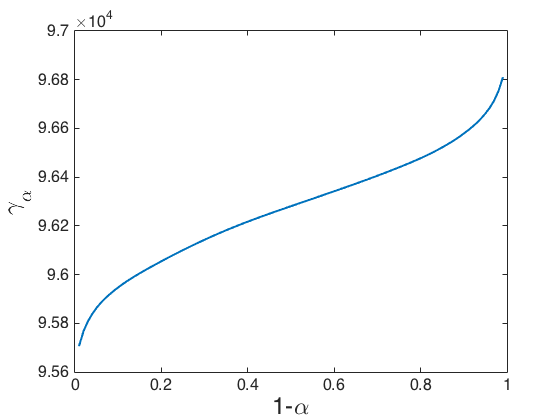}}
  \small{(a)}
\end{minipage}
\begin{minipage}[l2]{0.49\linewidth}
  \centering
  \centerline{\includegraphics[width=7.5cm]{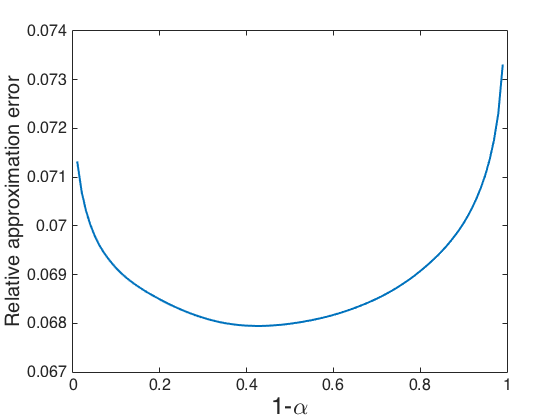}}
  \small{(b)}
\end{minipage}
\caption{\small{Microscopy experiment: (a) HDP region thresholds $\gamma_\alpha$, (b) relative approximation error $(\tilde{\gamma}_\alpha-\gamma_\alpha)/\gamma_\alpha$.}} \label{FigMicro2}
\end{figure}

Finally, we conclude this experiment by analysing the approximation errors related to using $\tilde{C}_{\alpha}$ instead of the exact HPD credibility region $C^*_\alpha$ for this problem. Figure \ref{FigMicro2}(a) shows the value of the exact threshold $\gamma_\alpha$ for different values of $\alpha$ (computing these threshold values by Monte Carlo integration with the algorithm \cite{Pereyra2015} required $24$ hours). Similarly to section \ref{tomographic_imaging}, we observe that the difference between the thresholds $\gamma_{0.01}$ and $\gamma_{0.99}$ is very small, confirming that that the posterior probability mass is highly concentrated as established in Lemma \ref{Lemma1}. Moreover, Figure \ref{FigMicro2}(b) reports the relative error $(\tilde{\gamma}_\alpha - \gamma_\alpha)/\gamma_\alpha$. We observe that in this case the approximation errors are of the order of $7\%$, which as mentioned previously is very low because $\tilde{C}_{\alpha}$ is a conservative approximation of ${C}^*_{\alpha}$ for the class of log-concave distributions (hence it is generally not tight for a specific model and dataset), and that the approximation is available as a by-product of MAP estimation with minimum computational cost.

\section{Conclusion}
\label{sec:conclusion}
This paper presented a new and general methodology to compute approximate credible regions for inverse problems that are convex and possibly very high-dimensional. These approximations were derived by using inequalities and concentration of measure results related to information theory for log-concave random vectors. The approximations have many important theoretical and computational properties. First, they are conservative regions that by construction contain the true high-posterior-density credible sets $C^*_\alpha$. As a result they can be used to reject point hypotheses of the form $\bx \in C^*_\alpha$ exactly. Second, their approximation error is bounded explicitly and, for large problems, grows at most linearly with the model dimension, comparing very favourably with other approaches that suffer from a curse of dimensionality. Moreover, from a computation viewpoint, the approximations can be calculated straightforwardly by using convex optimisation techniques that are several orders of magnitude faster than Bayesian computation methods based on Monte Carlo integration algorithms. In particular, the approximations are available as a free by-product in problems solved by maximum-a-posteriori estimation, which is currently the predominant Bayesian approach in mathematical imaging. Finally, the proposed approximations were illustrated with two mathematical imaging examples: tomographic image reconstruction with a total-variation prior, and sparse image deconvolution with an $\ell_1$ prior. In these examples the approximations were used to perform a range of point hypothesis tests and explore the uncertainty about specific aspects of the solutions delivered. To benchmark the approximations, the same analyses were conducted by using a state-of-the-art Monte Carlo algorithm that calculated the exact high-posterior-density credible regions. These comparisons showed that the approximations are remarkably accurate in spite of their simplicity.

As mentioned previously, because the approximations are conservative they can be used to reject point hypotheses with respect to $C^*_\alpha$ exactly. This is key in high-dimensional applications where the computation of $C^*_\alpha$ is challenging. Of course, the approximations can also be used to establish failure to reject hypotheses with an approximate confidence level. In some applications this approximation error is acceptable, either because the goal is to perform a coarse uncertainty analysis, or because the magnitude of the error can be characterised by conducting pilot experiments. Moreover, in sensitive applications the approximations can be used to as a preprocessing methodology to screen large datasets, followed by exact analyses by Monte Carlo integration for critical decisions and specific data.

Furthermore, this work opens many interesting perspectives for future research. For example, to investigate better approximations for subclasses of log-concave distributions (e.g., distributions that are strongly convex, or Lipschitz continuously differentiable), as well as inner-bounding approximations to complement the conservative approximation proposed in this work (however, in view of Theorem \ref{Theo2}, this will only be possible for specific subclasses of log-concave distributions). Another perspective is to develop computationally efficient algorithms specifically designed for approximating the posterior expectation $\textrm{E}\{g_{\by}(\bx)\}$; this would allow by-passing inequality \eqref{inequality1} and deriving significantly more accurate approximations based exclusively on the inequality of Lemma \ref{Lemma1}. {For example, one could consider a variational Bayesian approximation of $\textrm{E}\{g_{\by}(\bx)\}$.} Lastly, because statistical models are abstract representations that are inherently misspecified, it would be interesting to analyse the impact of model misspecification in uncertainty quantification and decision theory for mathematical imaging, both from Bayesian and frequentist statistical perspectives.

{Also, in this work we assume that the regularisation parameter $\lambda$ is fixed a-priori, and therefore do not take into account any potential posterior uncertainty stemming from it. Similarly, because $\lambda$ is specified by the practitioner, two analysts using significantly different values of $\lambda$ may potentially arrive to different conclusions about a same dataset. Consequently, another important perspective for future work is to extend the proposed methodology to cases where the value of $\lambda$ is fully or partially unknown and estimated from data, for instance by building on the hierarchical Bayesian approach \cite{Pereyra_EUSIPCO_2015} that estimates $\bx$ and $\lambda$ jointly. Precisely, evaluating $\tilde{C}_\alpha$ with the estimate $\hat{\lambda}$ of \cite{Pereyra_EUSIPCO_2015} produces an approximation of the marginal HPD of $\bx$ that is accurate when $p(\lambda|\by)$ is highly concentrated (e.g., when $\by$ is high-dimensional and a Bernstein-von Mises theorem holds for $p(\by|\lambda)$, enabling Laplace's method \cite{mackay}). This approach has a computational cost that is similar to the case where $\lambda$ is fixed. Otherwise, the variability of $p(\lambda|\by)$ could be incorporated by adopting a sampling approach such as Bayesian bootstrapping \cite{Laird87}, with a higher computational cost.}

{Moreover, as mentioned earlier, the experiments reported in Section \ref{sec:experiments} use $n$-dimensional credible regions to explore properties of subsets of pixels of dimension $m \ll n$. This approach has operational advantages, but it generally overestimates uncertainty \footnote{{Let $\overline{\bx} \in \mathbb{R}^m$ be the subset of pixels of interest, $\underline{\bx}\in \mathbb{R}^{n-m}$ the remaining pixels of $\bx$, $p(\overline{\bx}|\by) = \int p(\overline{\bx}, \underline{\bx}|\by) \textrm{d} \underline{\bx}$ the marginal posterior of interest, and $\bar{C}_\alpha$ the projection of $C^*_{\alpha}$ on the $m$-dimensional subspace of $\overline{\bx}$. It is easy to check that $\int_{\bar{C}_\alpha} p(\overline{\bx}|\by) \textrm{d} \overline{\bx} \geq 1-\alpha$, implying that the approach is generally conservative.}}. From a statistical viewpoint it would be more accurate to perform analyses with the marginal posterior $p(\overline{\bx}|\by) = \int p(\overline{\bx}, \underline{\bx}|\by) \textrm{d} \underline{\bx}$. Because $p(\bx|\by)$ is log-concave this marginal is also log-concave, i.e., $p(\overline{\bx}|\by) \propto \exp{\{-\overline{g}_{\by}({\overline{\bx}})\}}$ for some convex function $\overline{g}_{\by}$. Hence the proposed methodology could be applied to $p(\overline{\bx}|\by)$. Unfortunately, $p(\overline{\bx}|\by)$ and $\overline{g}_{\by}$ are generally computationally intractable and cannot be evaluated exactly. The development of accurate convex approximations of $\overline{g}_{\by}$ to use as surrogates is currently under investigation.}

Finally, although this work focused mainly on enabling the computation of credible regions for large-scale inverse problems, we hope and anticipate that future work will build on it to develop approaches for visualising and summarising credible regions, particularly in the context of hypothesis tests for imaging inverse problems. This will certainly contribute significantly to the progress of mathematical imaging methodology and to its capacity to support formal decision-making and scientific inquiry.

\section{Acknowledgements}
The author holds a Marie Curie Intra-European Research Fellowship for Career Development, and is grateful to Ben Powell, Jonty Rougier and Peter Green for useful discussion.
\bibliographystyle{plain}
\bibliography{refs}

\end{document}